\documentclass[twocolumn,showpacs,preprintnumbers,amsmath,amssymb]{revtex4}

\usepackage{graphicx}
\usepackage{dcolumn}
\usepackage{bm}

\begin{document}

\preprint{APS/123-QED}

\title{Aging phenomena in PMMA thin films\\
-- memory and rejuvenation effects}

\author{K. Fukao}
 \email{fukao@kit.ac.jp}
 \altaffiliation[]{Corresponding author.}
\author{A. Sakamoto}%
 \altaffiliation[Present address : ]{Department of Macromolecular
 Science, Graduate School of Science, Osaka University, Toyonaka, Osaka 560-0043, Japan.}
\affiliation{%
Department of Polymer Science, Kyoto Institute of Technology, 
Matsugasaki, Kyoto 606-8585, Japan
}%


\date{\today}

\begin{abstract}
Aging dynamics in thin films of poly(methyl methacrylate) (PMMA) have been 
investigated through dielectric measurements for different types of
aging processes.
The dielectric constant was found to decrease with increasing aging time at
an aging temperature in many cases. An increase in the dielectric
constant was also observed in the long time region ($\ge$11h) near the 
glass transition temperature for thin films with thickness less than 26nm.
In the constant rate mode including a temporary stop at a temperature
 $T_a$, the memory of the aging at $T_a$ was found to be kept and then
 to be recalled during the subsequent heating process.
In the negative temperature cycling process, a strong rejuvenation
effect has been observed after the temperature shift from the initial
temperature $T_1$ to the second temperature $T_2$($=T_1+\Delta T$) when
$\Delta T\approx -20$K. Furthermore, a full memory effect has also been
observed for the temperature shift from $T_2$ to $T_1$.
This suggests that the aging at $T_1$ is totally independent of that
at $T_2$ for $\Delta T\approx -20$K. 
As $|\Delta T|$ decreases, the independence of the aging 
between the two temperatures was found to be weaken, $i.e.,$ 
the effective time,
which is a measure of the contribution of the aging at $T_1$ to that at
$T_2$, is a decreasing function of $|\Delta T|$ in the negative region of
$\Delta T$. As the film thickness decreases from 514nm to 26nm, the 
$|\Delta T|$ dependence of the effective time was found to become 
much stronger. The contribution of the aging at $T_2$ to that at $T_1$ 
disappears more
 rapidly with increasing $|\Delta T|$ in thin film geometry than in the
 bulk state.
\end{abstract}


\maketitle

\section{Introduction}
 In glassy materials it is well-known that a very slow relaxation
 towards an 
equilibrium state is observed below the glass transition
temperature $T_g$~\cite{Struick,Bouchaud}. 
 This slow relaxation is called aging and is regarded as 
an important common
property characteristic of disordered materials such as spin
glasses~\cite{Lefloch,Vincent,Jonason1,Jonason2}, orientational
glasses~\cite{Doussineau}, supercooled liquids~\cite{Leheny}, relaxor
ferroelectrics~\cite{Kircher},  and polymer
glasses~\cite{Bellon1,Bellon2}. The investigations on such
disordered systems revealed that {\it memory and rejuvenation}
effects are observed during the aging process.  
The characteristic behavior of the aging dynamics below $T_g$ is
closely related to the nature of the glass transition. Therefore, it 
is expected
that the elucidation of the aging phenomena can lead to full understanding
of the mechanism of the glass transition in disordered materials, which
is still a controversial topics~\cite{IDMRCS4}.

In polymeric systems, Kovacs {\it et~al.} investigated the memory effects 
observed in the aging phenomena~\cite{Kovacs}. 
In their experiments, a volume relaxation was measured on 
a glassy state located on the line extrapolated from
the equilibrium line corresponding to a liquid state. Because the 
glassy state is on the
equilibrium line, it is expected that no further volume change
occurs on this state. However, their results suggest that the volume
relaxation strongly depends on the thermal history which the
polymer had experienced before arriving at this equilibrium state.
In other words, the thermal history can be memorized within the glassy
state of polymers.
Recent experiments by Miyamoto {\it et~al.} revealed that polymeric materials
can keep not only thermal history but also mechanical history in their
glassy states and the histories can be recalled upon the glass
transition, through measurements of tension in
rubbers~\cite{Miyamoto}.   

In spin glasses, interesting studies on the aging phenomena have been
done in recent years~\cite{Lefloch,Vincent,Jonason1}. 
In the experiments on the CdCr$_{1.7}$In$_{0.3}$S$_4$ insulating 
spin glass, a relaxation of the
imaginary part of the magnetic ac-susceptibility $\chi''$ was observed in a
temperature cycling. First, the sample is cooled down to a
temperature $T_1$ below $T_g$ and is kept at this temperature for a
period of 
$\tau_1$. Then the temperature is changed from $T_1$ to $T_1+\Delta T$
($\Delta T<0$) and kept for 
$\tau_2$, and after that the temperature is returned to $T_1$. It
was observed that upon cooling from $T_1$ to $T_1+\Delta T$, the system
{\it rejuvenates} and returns to a zero-age structure even after a long stay
at the higher temperature $T_1$. The interesting point is that the
system has a {\it perfect memory} of the past thermal history, {\it i.e.,}
when heated back to $T_1$, the susceptibility $\chi''$ agrees
well with the value that $\chi''$ had reached just at the end of the
first stage of the aging at $T_1$. 
As for polymeric systems, a similar measurement on the aging dynamics has
been done for the dielectric constant in poly(methyl methacrylate)
(PMMA) by Bellon {\it et~al.} to
investigate the memory effects of the aging dynamics. Although a 
memory effect similar to that in spin glasses has been observed also in 
PMMA in their studies~\cite{Bellon1,Bellon2}, more detailed
investigations would be required to extract an universal behavior of the
aging dynamics in polymer glasses and to compare the aging dynamics
between polymeric systems and spin glasses. 

It is very natural to believe that the
aging dynamics in the glassy state mentioned above are related 
to the nature of the
glass transition. Recent investigations on the glass transition in thin
film geometry show that there is a strong dependence of the glass
transition temperature and the dynamics on film thickness and the system
size~\cite{Keddie,Fukao1,Confit2003}. Therefore, it is expected that the aging
dynamics depend on the film thickness~\cite{Kawana1}, and through
such investigations it would be possible to approach the basic mechanism
of the glass transition. From the technical point of view, thin film
geometry is ideal for dielectric measurements because the smaller
thickness of the films enhances the resolution of capacitance measurements.
For this reason, precise measurements can be done for thin film geometry.

In this study, we measured time evolution of the complex electric
capacitance of PMMA films during the aging process using dielectric 
relaxation spectroscopy.
The purpose of this study is to
investigate the relaxation dynamics of non-equilibrium states including the
memory and rejuvenation effects in polymeric systems and to elucidate
the thickness dependence of the aging dynamics.

This paper consists of six sections.
After giving experimental details in Sec.II, the aging dynamics at a
given temperature are shown in Sec.III. In Sec.IV, experimental
results on the memory and rejuvenation effects are presented for two
different thermal histories. After discussing the experimental results
in Sec.V, a summary of this paper is given in Sec.VI.

\section{Experiment}
Polymer samples used in this study are atactic
PMMA purchased from Scientific Polymer Products, Inc. 
The weight-averaged molecular
weight $M_{\rm w}$=3.56$\times$10$^5$ and $M_{\rm w}/M_{\rm n}$=1.07,
where $M_{\rm n}$ is the number-averaged molecular weight.
The glass transition temperature $T_{\rm g}$ determined by differential
scanning calorimetry (DSC)  
is about 380 K for the bulk samples of PMMA used in this study.
The value of $T_{\rm g}$ in thin films with $d$=26nm is lower by 7K
than $T_{\rm g}$ in the bulk samples. 
Thin films are prepared on aluminum vacuum-deposited glass
substrate using a spin coat method from a toluene solution of PMMA.
The thickness of the films is controlled to be 20nm to
514nm by changing the concentration of the solution of PMMA. The
thickness is evaluated from the value of the electric capacitance at 273 K 
of as prepared films before measurements in the same way as in 
previous papers~\cite{Fukao1,Fukao3,Fukao4}. 
After annealing at 343K, aluminum is vacuum deposited
again to serve as upper electrodes.
A copper wire is attached to the end of the electrode with
conducting paste so that the polymer samples can easily be
connected with the measurement system. 
The sample prepared in the above way is mounted inside a sample cell.
The temperature of the sample cell is controlled through a heater wound 
around the cell using a temperature controller. The temperature of 
the polymer sample is monitored 
as the temperature measured through a thermocouple put onto the back side of
the glass substrate. Heating above 400 K was done several times before 
the measurements so that the contact between the polymer films and 
the substrate is stabilized.   
The temperature of the sample is stabilized within 2 minutes after
a temperature jump. The
stability of the temperature control is $\pm$0.1K.

Dielectric measurements were done by 
an LCR meter (HP4284A). The frequency range of the applied electric field was
from 20Hz to 1MHz. The voltage level was controlled so that the strength
of the applied electric field remains almost constant independent of film 
thickness.
The typical electric field was 2$\times$10$^6$ V/m.
In our measurements, the complex electric capacitance of the sample
condenser $C^*(T)$ was measured and then converted into the dynamic (complex)
dielectric constant $\epsilon^*(T)$ by dividing the $C^*(T)$ by the 
geometrical capacitance $C_0(T_0)$ at a standard temperature $T_0$.
The value of  $C^*$ is given by $C^*=\epsilon^*\epsilon_0\frac{S}{d}$ and
$C_0=\epsilon_0 \frac{S}{d}$, 
where $\epsilon_0$ is the permittivity in vacuum, $S$ is the area of 
the electrode and $d$ 
is the film thickness. For evaluation of $\epsilon^*$ and $C_0$, 
we use the thickness $d$ which is determined at $T_0=$273 K and 
$S$=8$\times$10$^{-6}$mm$^2$.
Because $\epsilon^*$ and $C^*$ are complex numbers, we define the real
and imaginary parts as follows: $\epsilon^*$=$\epsilon'-i\epsilon''$ 
and $C^*$=$C'-iC''$.

\section{Aging Dynamics}
In this section, we show the relaxation behavior during the isothermal
aging process at a given temperature $T_a$.
In order to remove the thermal history of the sample, the system is 
heated up to 403K ($>T_g$) and is kept at 403K for a couple of 
hours. After this procedure the system is cooled down to $T_a$ at 
the rate of 0.5 K/min and the measurement of
the complex electric capacitance is started. Figure 1(a) displays the time 
evolution of the imaginary component of the dynamic dielectric constant
$\epsilon''$ at various aging temperatures for the thin films with
thickness of 20nm. In Fig.1(a), it is found that $\epsilon''$ is a
monotonical decreasing function of the aging time $t_w$. 
The vertical axis in this figure is the value of $\epsilon''$ normalized 
with respect to the initial value $\epsilon''(0)$. The origin of 
the time axis is defined as the time at which the temperature of the system
reached $T_a$. 
The decrease in the dielectric constant corresponds to the smaller
response against an applied external field, and hence this indicates that
the system goes down to a deeper valley of the free energy via thermal
activations within the energy landscape picture~\cite{Angell1,Debenedetti1}. 
In other words, the
system approaches an equilibrium state as the time $t_w$ elapses at a 
temperature $T_a$ below $T_g$.
The aging phenomena can be interpreted as this approach to the 
equilibrium state. 

\begin{figure}
\includegraphics*[width=8cm]{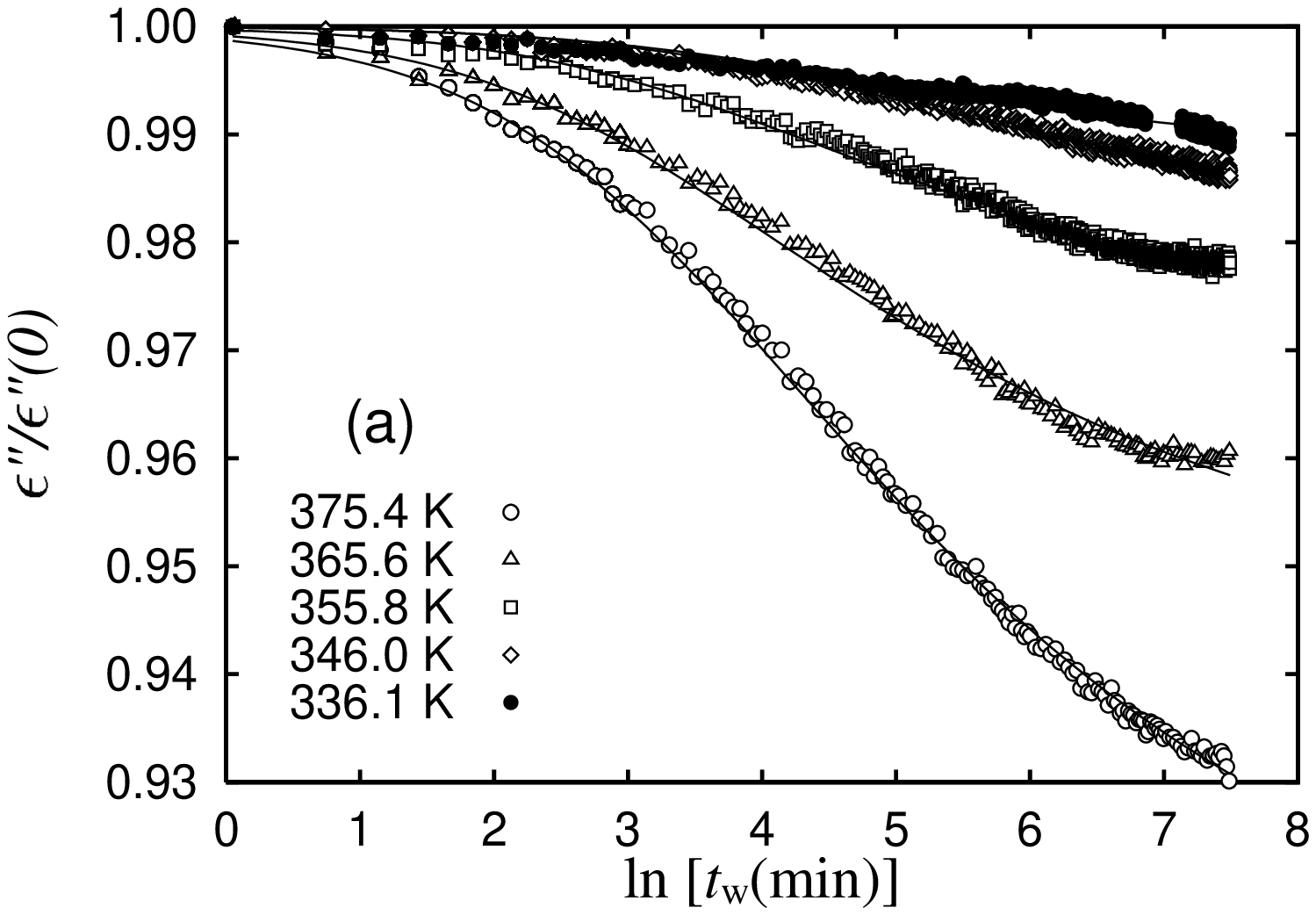}

\includegraphics*[width=8cm]{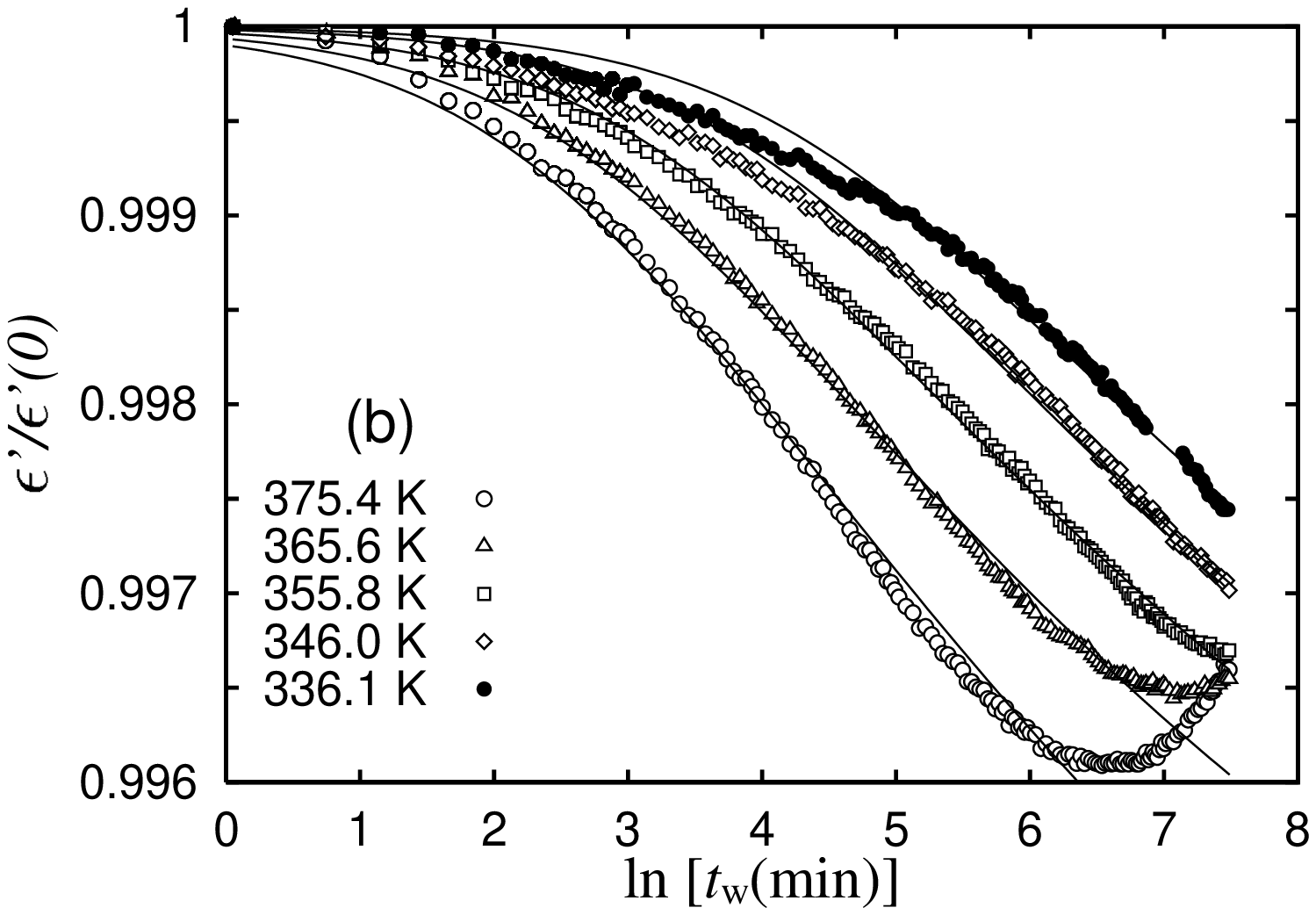}
\caption{\label{fig:1} Aging time dependence of the dynamic dielectric constant 
 normalized with the value at $t_w$=0 in PMMA thin films with $d$=20nm
 for various aging temperatures: $T_a$=375.4 K, 365.6 K, 355.8 K, 346.0
 K, and 336.1 K: (a) the imaginary component and (b) the real component. 
The frequency of the applied electric field $f$ is 20Hz. The horizontal 
axis is the natural logarithm of $t_w$. 
The solid curves are obtained by fitting the data
 points with Eq.(5). The best-fit parameters for the imaginary parts 
are listed in Table~I.}
\end{figure}

\begin{table}
\caption{\label{tab:table1}
Fitting parameters obtained by fitting the observed values of
 $\epsilon''$ with Eq.(5) for thin films of PMMA with
 $d=20$nm. The frequency of applied electric
 field is 20Hz. The exponent $n$ is fixed to be 0.309, which is obtained
 by fitting the data at 355.8 K}
\begin{ruledtabular}
\begin{tabular}{ccccc}
$T_a$ (K) & $\Delta\epsilon''_{ag}/\epsilon''(0)(\times 10^{-2})$
& $t_0$ (min) & $\epsilon''(0)$ & $\Delta\epsilon''_{ag}$ \\
\hline
375.4 & $9.286\pm0.007$ & 21.8$\pm$0.1  & 0.1147  & 0.01065 \\
365.6  & $5.486\pm0.007$ & 18.6$\pm$0.2  & 0.1302  & 0.00714 \\
355.8  & $3.081\pm0.041$ & 26.4$\pm$0.9  & 0.1601  & 0.00493 \\
346.0  & $1.989\pm0.006$ & 52.3$\pm$0.7  & 0.2016  & 0.00401 \\
336.1  & $1.357\pm0.007$ & 35.3$\pm$0.9  & 0.2495  & 0.00339 \\
\end{tabular}
\end{ruledtabular}
\end{table}

In an equilibrium state, the dielectric constant can be uniquely
specified as functions of temperature $T$, pressure $p$ and the 
frequency of the applied electric field $f$ for the polymer films 
with a fixed thickness. For simplicity, we ignore the effect of the 
pressure in the discussions below. 
If the system is in a non-equilibrium state, it is not easy to specify 
the dielectric constant,  and the aging
phenomena must be taken into account. The dielectric constant depends 
on the thermal history that the sample has experienced.
In general, if the system 
is aged at a temperature $T_{a,1}$ for a period of
$t_{w,1}$, at $T_{a,2}$ for $t_{w,2}$, $\cdots$, and 
at $T_{a,n}$ for $t_{w,n}$,
the dielectric loss $\epsilon''$ can be specified by
the following description :
\begin{eqnarray}
 \epsilon''=\epsilon''(T,\omega; \{t_{w,j}(T_{a,j})\}),
\end{eqnarray}
where $\omega=2\pi f$ and $\{t_{w,j}(T_{a,j})\}$ $\equiv$
$\{t_{w,1}(T_{a,1}), t_{w,2}(T_{a,2}),$  $\cdots$ $t_{w,n}(T_{a,n})\}$.
If the system is aged for $t_{w}$ at just one temperature 
$T_a$, the dielectric loss
at the temperature $T$ is described by 
\begin{eqnarray}
 \epsilon''=\epsilon''(T,\omega; t_{w}(T_{a})).
\end{eqnarray}
Firstly, the aging dynamics under an isothermal condition at
$T$=$T_a$ are considered. 
In principle, the dielectric loss $\epsilon''(T_a,\omega; \infty)$ 
is the value of $\epsilon''$ in the equilibrium state at $T_a$ and the 
rest of $\epsilon''$ can be defined as the aging part 
$\epsilon''_{ag}(T_a,\omega; t_w)$ in the following way:
\begin{eqnarray}\label{eps2}
\epsilon''(T_a, \omega, t_w)=\epsilon''(T_a,\omega; \infty)+
\epsilon''_{ag}(T_a, \omega; t_w).
\end{eqnarray}
The symbols $T_a$ and $\omega$, here, are dropped for simplicity and it
is assumed  
that the aging part of $\epsilon''$ is given by the following equation:
\begin{eqnarray}\label{e_ag}
\epsilon''_{ag}(t_w)=\frac{\Delta\epsilon''_{ag}}{(1+t_w/t_0)^n},
\end{eqnarray}
where $\Delta\epsilon''_{ag}$ is the relaxation strength towards the
equilibrium value, $t_0$ is the characteristic time of the aging
dynamics, and $n$ is an exponent. If $t_w$ is much larger than $t_0$, 
$\epsilon''_{ag}\approx t_w^{-n}$. In this case, the dielectric loss  
$\epsilon''$ decays according to the power-law decay in the long time
region. The similar expressions for $\epsilon'$ are also used in this paper. 

The solid curves given in Fig.1(a)
were obtained by a non-linear least square fit of the observed data to
the following equation derived from Eqs.(3) and (4):
\begin{eqnarray}\label{fit_eq}
\frac{\epsilon''(t_w)}{\epsilon''(0)}=1-\frac{\Delta\epsilon''_{ag}}
{\epsilon''(0)}\left[1-\left(1+\frac{t_w}{t_0}\right)^{-n}\right].
\end{eqnarray}
Table I displays the fitting parameters obtained by Eq.(\ref{fit_eq}) for 
$\epsilon''$ in thin films of PMMA with $d$=20nm at various 
aging temperatures, where the value of $n$ is fixed to be 
the value obtained for 355.8 K. 
The relaxation strength $\Delta\epsilon''_{ag}$ was found to decrease with 
decreasing aging temperature $T_a$. Furthermore, the characteristic
time $t_0$ of the relaxation of $\epsilon''$ in the aging process
was found to increase with decreasing $T_a$, in other words, the aging 
dynamics become slower with decreasing $T_a$. Because the
observed data in Fig.1(a) can well be reproduced using Eq.(4), it is
implied that the relaxation of $\epsilon''$ at an isothermal condition 
below $T_g$ obeys the power-law one for the long-time region in the
thin films with $d=20$nm. 
According to the report in Ref.~\cite{Bellon1}, the aging 
behavior depends on the frequency of the applied electric 
field. As the frequency decreases, the relaxation strength due to
aging increases for the frequency range from 0.2Hz to 20Hz. In this 
study, we have measured the dynamic dielectric constant over 4 decades
and could confirm the results in Ref.~\cite{Bellon1}.


As mentioned later, under the present experimental conditions that  
$T_a=336.1$K$\sim 375.4$K and $f=20$Hz are located between the 
$\alpha$-process and the $\beta$-process in the dispersion map. Because
$\Delta\epsilon''_{ag}$ decreases with decreasing $T_a$,
the relaxation of $\epsilon''$ due to the aging may be associated with
the $\alpha$-process. 

Figure 1(b) displays the $t_w$ dependence of the real part of the
complex dielectric constant $\epsilon'$ normalized with respect to 
the initial value at various aging temperatures $T_a$. 
In this figure, it is found that $\epsilon'$ decreases with increasing 
$t_w$ in a similar way to $\epsilon''$ for $T_a$'s less than 365.6 K.
However, there is a surprising difference between $\epsilon'$ 
and $\epsilon''$ for $T_a$=375 K: $\epsilon'$ decreases with $t_w$ up to 
$t_w\approx$ 11h and then begins to increase with $t_w$. This increase in
$\epsilon'$ in the long time region was not observed for lower values of
$T_a$. A similar increase in $\epsilon'$ was observed also for $d=$26 nm. 
On the contrary, for thicker films ($d=$78, 514 nm), no such increase 
in $\epsilon'$ was observed in the present measurements. 

\begin{figure}
\includegraphics*[width=8cm]{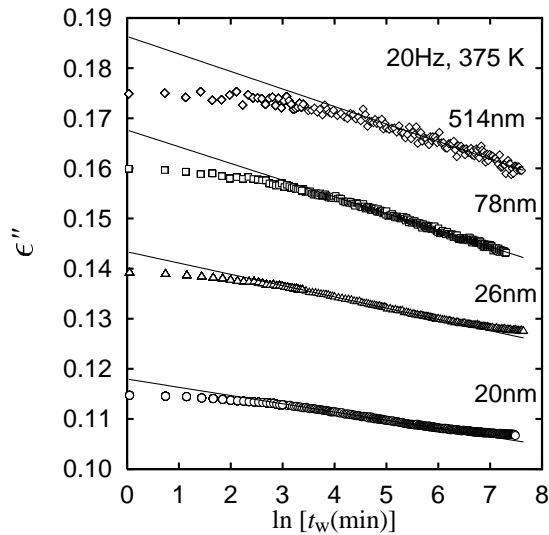}
\caption{\label{fig:2} Aging time dependence of the dielectric loss as a function of film thickness
 for PMMA thin films: $d$=20nm, 26nm, 78nm and 514nm, $T_a$=375K and
 $f$=20Hz. A deviation from the straight line implies that the logarithmic
 decay does not hold in the long time region for $d$=20nm and 26nm.
}
\end{figure}

We here move to the results for the thin films with various film
thickness to see the thickness dependence of the aging dynamics.
Figure 2 displays the time variation of the dielectric loss $\epsilon''$ 
at $T_a=$375 K for the thin films with thickness between 20nm and 514nm. 
Because the horizontal axis in Fig.2 is on a logarithmic scale, any 
straight line indicates the existence of a logarithmic decay of the 
dielectric loss. For $d$=20nm and 26nm, we can see that 
there is a slight deviation from the straight line in the short and 
long time regions.
This is consistent with the result that the time variation in
$\epsilon''$ obeys Eq.(5) for $d$=20 nm, as shown in Fig.1(a).
For $d=$78nm and  514nm, on the other hand, $\epsilon''$
decreases linearly with $\log t_w$ for the region of $t_w > 20$
min. Within the present time window there is no tendency to
saturate to a constant value. In this case, for $t_w>20$min, the 
relaxation of $\epsilon''$ due to aging obeys a logarithmic law. This
logarithmic decay law was observed also in the measurements done by 
Bellon {\it et~al.}~\cite{Bellon1}. 
Although it is impossible to judge whether the logarithmic
decay is valid also outside the time window, there may be
a qualitative difference between the decay law for the thin films and that 
for the thicker films.

In Sec.III, we have observed that there is the decrease in $\epsilon'$ 
and $\epsilon''$ with increasing $t_w$ at various values of 
$T_a(=336.1$K$\sim 375.4$K), 
except under the condition that $t_w>$ 11h, $d\le $26nm, and
$T_a$=374K, where the increase in $\epsilon'$ with $t_w$ was observed. 
In the discussion  section, the physical origin of the time
variation of $\epsilon'$ and $\epsilon''$ will be discussed.


\section{Memory and rejuvenation effects}

In Sec.IV we show the memory and rejuvenation effects observed during
the aging process and discuss the detailed properties including the thickness 
dependence. For this purpose, we adopt two modes of thermal 
history: 1) 
{\it constant rate mode} (CR mode) -- After the thermal equilibration at
403 K, the sample is cooled at a rate of 0.5
K/min down to a temperature $T_a$, and then is
isothermally annealed (aged) at $T_a$ for 10 hours. 
The sample is then cooled down to room temperature at 0.5 K/min. 
Finally, the sample is heated again from room temperature 
to 403 K at 0.5 K/min.
2) {\it temperature cycling mode} (TC mode) -- The sample is cooled
at a rate of 0.5 K/min from 403 K down to a temperature 
$T_1$ and is aged at $T_1$ for a period  of  
$\tau_1$. Then, the temperature is rapidly changed
from $T_1$ to $T_2$ ($T_2\equiv T_1+\Delta T$) and is kept at
$T_2$ for $\tau_2$. Finally, the temperature is changed back from $T_2$
to $T_1$ and is kept at $T_1$ for $\tau_3$. In this section, we
show the data for $\tau_1=\tau_2=5$h. Both the modes
have originally been developed in order to investigate the relaxation
behavior in the aging phenomena in spin 
glasses~\cite{Lefloch,Vincent,Jonason1}.

\subsection{Constant rate mode}
As it is well-known, there are two typical molecular motions in PMMA, which
are usually called the $\alpha$-process and the
$\beta$-process~\cite{McCrum1}.  The $\alpha$-process is strongly
related to the glass transition, while the $\beta$-process is due to the
local motion of polymer chains. In the case of PMMA, the dielectric
relaxation strength of the $\beta$-process is larger than that of the
$\alpha$-process due to the existence of polar bulky side groups. Both the
dielectric constant and loss have a strong temperature dependence due to
the $\alpha$- and $\beta$-processes. For example, $\epsilon''$ exhibits 
two distinct loss peaks in a temperature domain at a given frequency :  
one is at about 310 K due to
the $\beta$-process and the other is at about 400 K  due to the
$\alpha$-process for $f=20$Hz and $d=20$nm. (See the inset of 
Fig.6.) 

For this CR mode, we measured both the dielectric constant 
$\epsilon'$ and the dielectric loss $\epsilon''$.  If the range 
of $t_w$ is restricted to $t_w<$ 10 hours, there is no qualitative 
difference in the aging dynamics between $\epsilon'$ and $\epsilon''$. In 
the case of $d$=514nm, the errors attached to the value of $\epsilon''$ 
are larger than those of $\epsilon'$ because of  the resolution of 
the LCR meter. For this reason, we show here the results of $\epsilon'$,
which can be described as $\epsilon'(T,\omega; t_w(T_a))$ according 
to our notation. 
In addition to the $\epsilon'$, we prepare a reference curve, 
$\epsilon'(T,\omega; 0)$, 
both for the heating and cooling processes at the same rate (0.5K/min) 
{\it without any temporary stop during cooling process}. 
We refer to the curve as $\epsilon'_{\rm
ref}$($\equiv\epsilon'(T,\omega; 0)$). 

\begin{figure}
\includegraphics*[width=8cm]{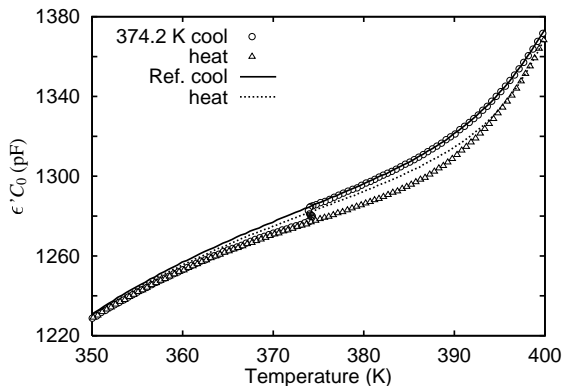}
\caption{\label{fig:3} Temperature dependence of the real part of the complex electric
capacitance $\epsilon'C_0$ observed by the CR mode with $T_a=374.2$K
for PMMA films with $d$=514nm. The frequency of the applied electric
field $f$ is 20 Hz. The symbols $\circ$ and $\triangle$ stand for 
the cooling and heating processes of the CR mode, 
respectively. The solid curve ($-$) and dotted curve ($\cdots$) are
for the reference data observed during the cooling and heating processes 
without any temporary stop at $T_a$.
} 
\end{figure}

Figure 3 shows an example of the temperature dependence of the 
real part of the complex electric capacitance $\epsilon'C_0$ 
observed by the CR mode with $T_a$=374.2 K for the films with
$d$=514nm. The reference curves $\epsilon'_{\rm ref}C_0$ for the
cooling process and the subsequent heating process are given 
by the solid and dotted curves, respectively.
The curve of $\epsilon'_{\rm ref}$ during the cooling process 
is slightly different from that during the heating process.
The data points denoted by circles stand for $\epsilon'C_0$ observed
during the cooling process, while those denoted by triangles do for
$\epsilon'C_0$ observed during the heating process.

\begin{figure}
\includegraphics*[width=8cm]{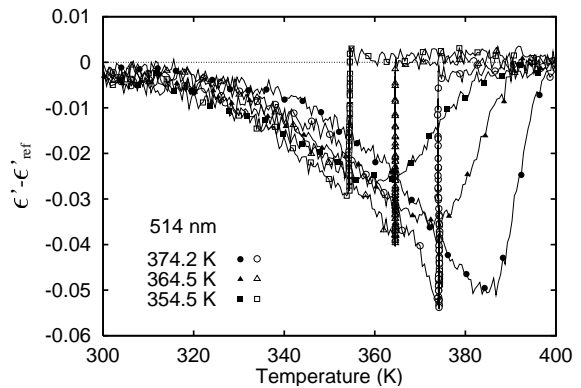}
\caption{\label{fig:4} Temperature dependence of the difference between $\epsilon'$ and
$\epsilon'_{\rm ref}$ observed by CR mode with three
 different aging temperatures $T_a$=374.2 K, 364.5 K and 354.5 K for
 PMMA films with $d$=514nm. The frequency of the applied electric field $f$
 is 20Hz. The heating and
 cooling rates were 0.5 K/min and the aging times at $T_a$ was 10
 hours. Open and full symbols display the data observed during 
the cooling and heating processes, respectively.
} 
\end{figure}

Figure 4 displays the temperature change in $\epsilon'-\epsilon'_{\rm ref}$, 
{\it i.e.}, the
difference between $\epsilon'$ and $\epsilon'_{\rm ref}$ obtained by
the CR mode with various aging temperatures $T_a$ for $d=514$nm.  
Here, we use
$\epsilon'_{\rm ref}$ of the cooling (heating) process for $\epsilon'$ of
the cooling (heating) process.
In Fig.4, it is found that, as the sample is cooled down from 403 K 
to $T_a$ (=374.2K), 
$\epsilon'$ is almost equal to $\epsilon'_{\rm ref}$ and the difference
$\epsilon'-\epsilon'_{\rm ref}$
remains zero. 
The difference then begins to deviate from 0 to a negative
value and the deviation becomes larger  monotonically with 
increasing aging time (at least up to $t_w\sim$ 10 hours)
during the subsequent isothermal aging at $T_a$. 
This indicates that
$\epsilon'$ decreases with aging time at $T_a$.
As the temperature decreases from $T_a$ to room temperature after
the isothermal aging, the
difference $\epsilon'-\epsilon'_{\rm ref}$ decreases with
decreasing temperature and becomes almost zero at room temperature. 
Just after
the temperature reaches 293 K, the heating process at the rate of 0.5K/min
starts. As the temperature increases, the difference
$\epsilon'-\epsilon'_{\rm ref}$ 
begins to deviate from 0 to a negative value almost along the same way
as observed during the preceding cooling process. The difference 
between $\epsilon'$ and $\epsilon'_{\rm ref}$ exhibits a maximum 
at around $T_a+$10K
and then decreases to 0 with increasing temperature. 

\begin{figure}
\includegraphics*[width=8cm]{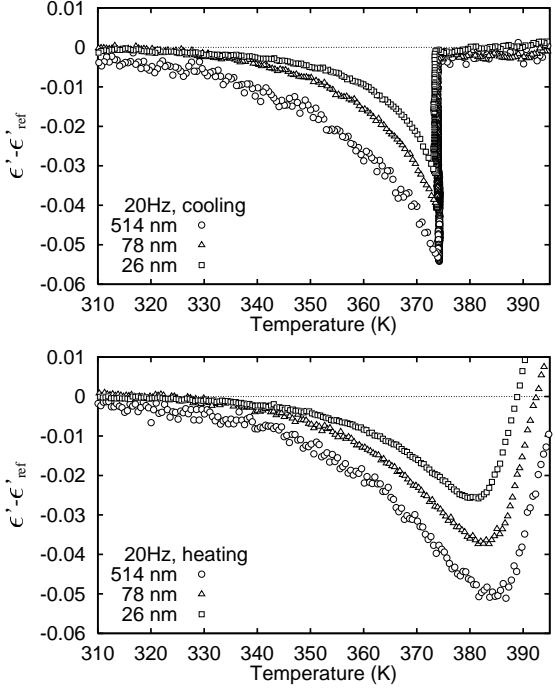}
\caption{\label{fig:5} Difference between $\epsilon'$ and
 $\epsilon'_{\rm ref}$ as functions of temperature observed by the CR
 mode for PMMA thin films with $d$=26nm, 78nm, and 514nm. The upper
 figure displays the results for the cooling process and the lower one does
 those for the subsequent heating process. Dielectric measurements were
 done for $f$=20Hz. The isothermal aging was done at $T_a=374$K during
 the cooling process.
}
\end{figure}

This behavior observed by the CR mode can be interpreted as follows: 
the thermal history that the sample is aged at $T_a$ for 10 hours is 
memorized at $T_a$ during the cooling process. As the
temperature decreases from $T_a$ to room temperature, the sample begins
to rejuvenate and is back to almost a standard age at room
temperature. During the subsequent heating process, the sample
becomes older almost according to the curve along which the sample 
experienced
rejuvenation during the preceding cooling process after a temporary stop at
$T_a$. This result implies that not only the aging process at $T_a$ 
near $T_g$ but also the subsequent cooling process can be memorized
and the whole thermal history can be read out during the heating process.
A similar temperature change in $\epsilon'-\epsilon'_{\rm ref}$ has 
already been observed in PMMA films of thickness of 0.3 mm~\cite{Bellon1}.

To investigate the thickness dependence of this aging behavior, we
show the results for various film thicknesses, $d=$26, 78, and 514nm. 
In Fig.5, it is found that
the strength of aging for 10 hours at 374 K decreases with
decreasing film thickness. Furthermore, during the subsequent cooling
process, the difference between $\epsilon'$ and $\epsilon'_{\rm ref}$
approaches zero more quickly in the thin films than in the thicker films. In
other words, the thin films are found to age more slowly, but to 
rejuvenate more quickly.

\begin{figure}
\includegraphics*[width=8cm]{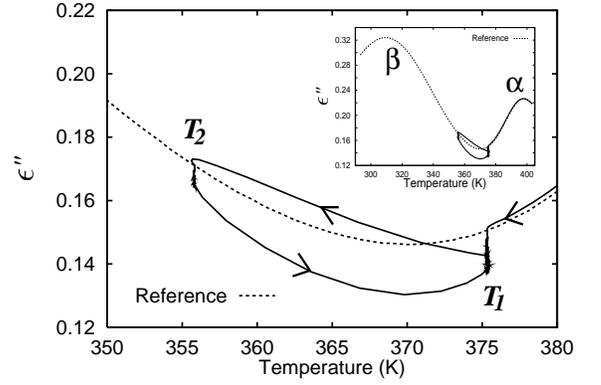}
\caption{\label{fig:6} Temperature dependence of the dielectric loss
 observed by the TC mode with $T_1$= 375.3K and $T_2$= 355.8K 
($\Delta T=-19.5$K) for PMMA thin films with $d$=20nm and the frequency
 of the applied electric field $f$=20Hz. The dotted curve displays 
the temperature
 dependence of $\epsilon''$ obtained during the cooling process with the
 rate of 0.5K/min without any isothermal aging. The overall behavior of
 $\epsilon''$ for temperature range from 310K to 403K is shown in the inset.
}
\end{figure}

\subsection{Temperature cycling with negative $\Delta T$}

In this subsection, we show the results obtained by the TC mode. 
In this mode, the sample is aged at two different temperatures
$T_1$ and $T_2$, and hence $\epsilon''$ is described as the
quantity $\epsilon''(T,\omega ; t_{w1}, t_{w2})$, where $t_{w1}$ and
$t_{w2}$ are the aging times at temperatures $T_1$ and $T_2$,
respectively. Here, the total aging time $t_w$ is given by 
$t_w=t_{w1}+t_{w2}$.
Using this expression, the dielectric loss $\epsilon''$ observed by 
the TC mode can be described in the following way:
\begin{eqnarray}\label{eps2_age}
\epsilon''
= \left\{ \begin{array}{c@{\quad:\quad}l} 
\epsilon''(T_1,\omega ; t_w,0)  & 0<t_w<\tau_1 \\ 
\epsilon''(T_2,\omega ; \tau_1, t_w-\tau_1)  & \tau_1<t_w<\tau_1+\tau_2 \\ 
\epsilon''(T_1,\omega ; t_w-\tau_2, \tau_2)  & t_w > \tau_1+\tau_2. 
\end{array}\right. 
\end{eqnarray}
For example, for a value of $t_w$ ($\tau_1<t_w<\tau_1+\tau_2$),
the notation $\epsilon ''(T_2,\omega ; \tau_1,t_w-\tau_1)$ 
corresponds to the imaginary part of the dielectric constant 
observed by frequency $f=\omega /2\pi$ at $T_2$ 
after the aging at $T_1$ for $\tau_1$ and at $T_2$ for $t_w-\tau_1$. 
In this case, the reference curve $\epsilon''_{\rm ref}$ can
also be defined as $\epsilon''_{\rm ref}=\epsilon''(T,\omega ;
0,0)$.

\begin{figure*}
\includegraphics*[width=8cm]{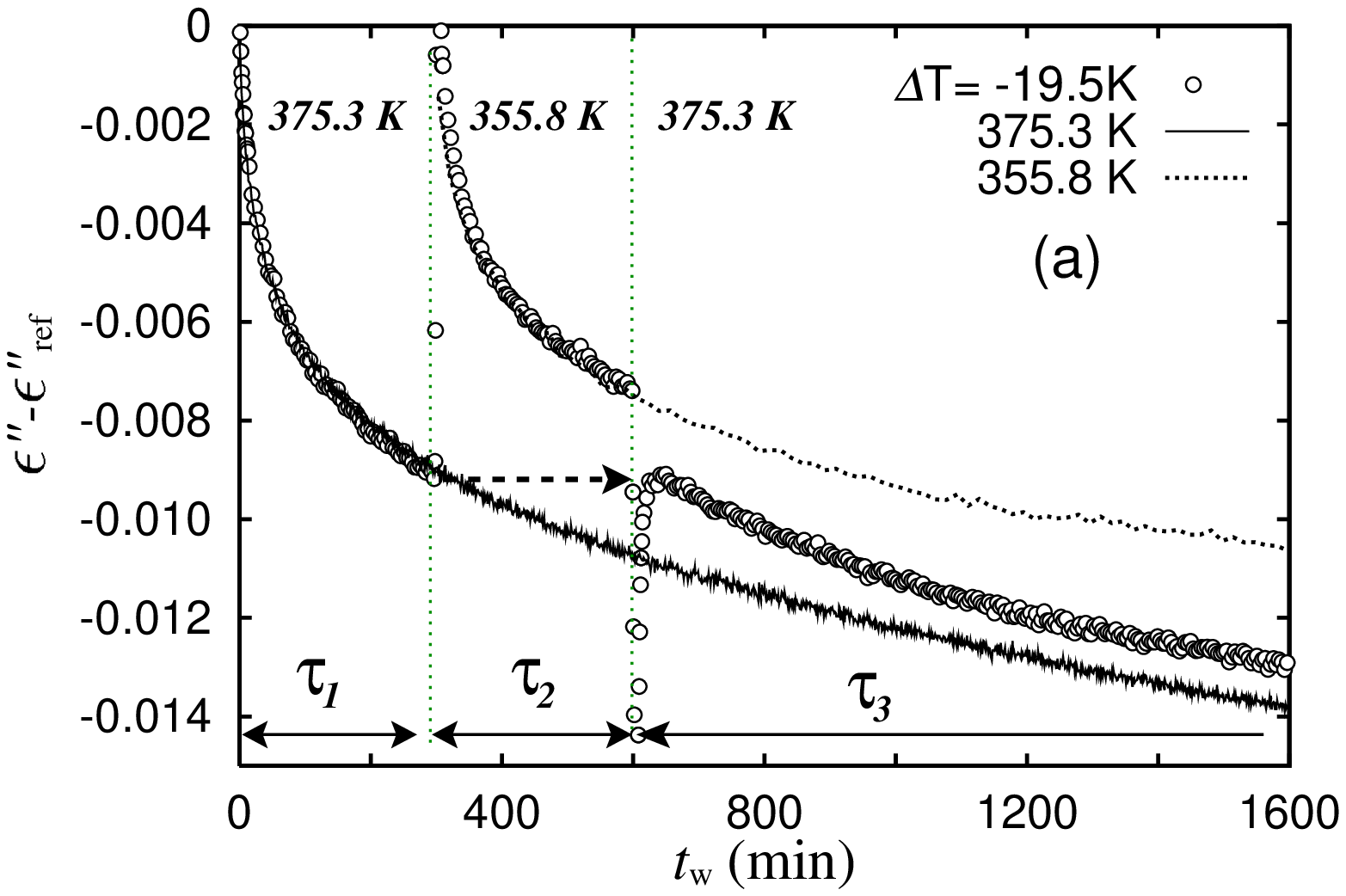}
\includegraphics*[width=8cm]{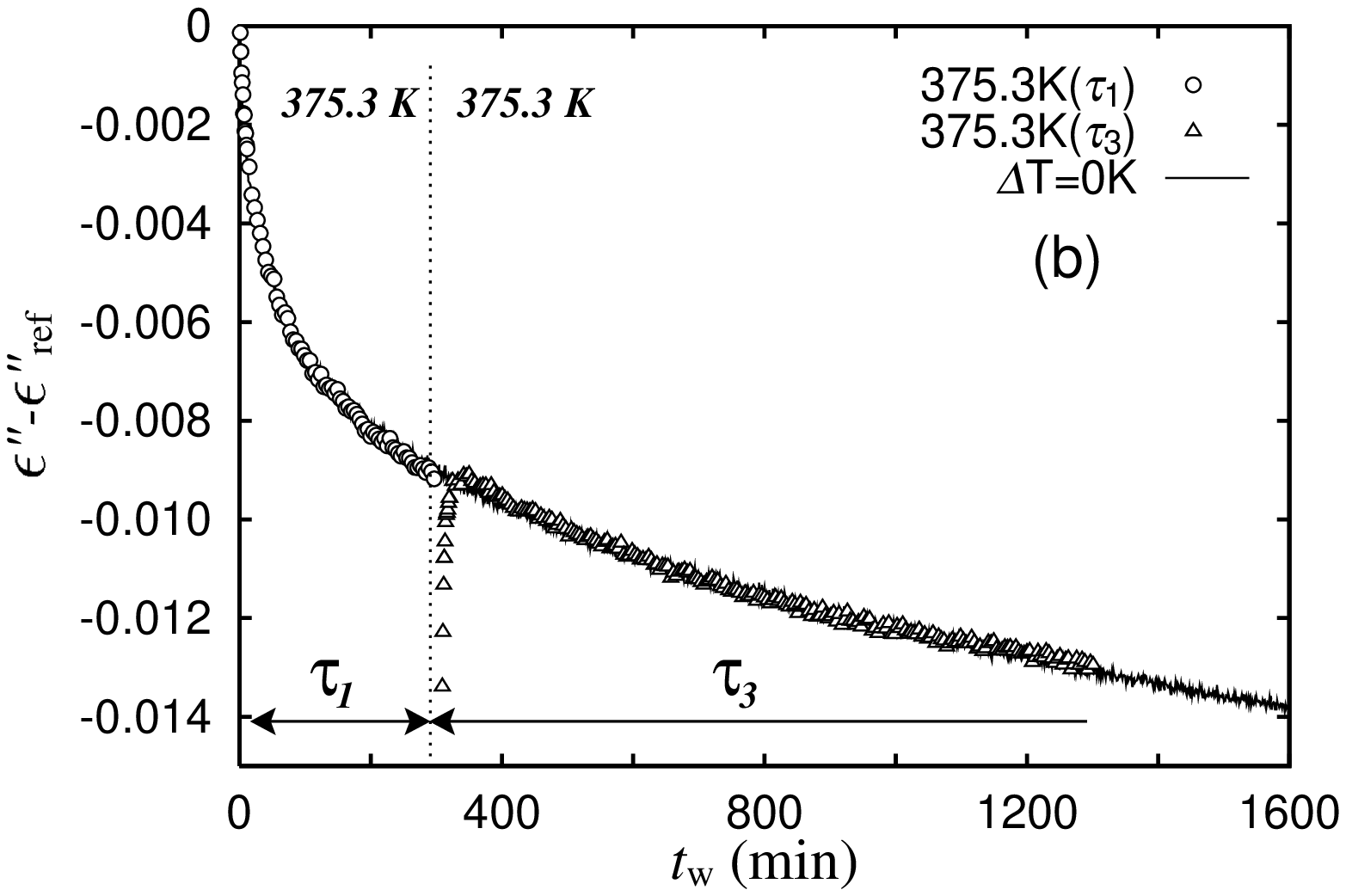}
\caption{\label{fig:7} (a) Difference between $\epsilon''$ and
 $\epsilon''_{\rm ref}$ observed by the TC mode with $T_1$=375.3K and
 $T_2$=355.8K ($\Delta T$=$-$19.5K) for PMMA thin films with $d$=20nm. 
(b) The difference $\epsilon''-\epsilon''_{\rm ref}$ obtained by shifting the
 date points in the third stage in the negative direction of the time
 axis by $\tau_2$ after removing the data points in the second
 stage.  Aging times at the first and second stages are
 $\tau_1=\tau_2=5$ hours. 
The horizontal axis of Fig.7(b) is the total aging 
time at $T_1$. The solid ($-$) and dotted curves ($\cdots$) are 
standard relaxation ones obtained by isothermal aging at $T_1$=375.3 K 
and $T_2$=355.8 K, respectively. It should be noted that the time origin of the
dotted curve for $T_2$ is shifted from $t_w$=0 to $t_w$=$\tau_1$.
}
\end{figure*}

Figure 6 displays a typical behavior of $\epsilon''$ and $\epsilon''_{\rm ref}$ 
of the TC mode for PMMA thin films with $d=20$nm and $f=20$Hz. In the
inset, the temperature dependence of $\epsilon''_{\rm ref}$ at 20Hz 
is shown for
the temperature range from 290K to 403K during the cooling process (See
the dotted curve). The two distinct loss peaks, the $\alpha$-process and
the $\beta$-process, are observed. 
As the sample is cooled from 403K to $T_1$(=375.3K) 
at the rate of 0.5K/min according to the procedure of the TC mode, 
the dielectric loss $\epsilon''$ moves 
along the reference curve $\epsilon''_{\rm ref}$ and then deviates 
from the reference curve to a lower value upon the beginning of the aging at
$T_1$ (See the solid curve). The main part of Fig.6 shows this 
behavior in the temperature range from 350 K
to 380 K. As the temperature decreases from $T_1$ to $T_2$(=355.8K) by
$\Delta T$=$-$19.5 K after the aging at $T_1$ for a period of 
$\tau_1$, the dielectric loss $\epsilon''$ immediately returns 
to a value on the reference curve
at $T_2$. As the temperature is kept at $T_2$, $\epsilon''$
deviates from the value on the reference curve. 
While the temperature is changed from $T_2$ to $T_1$ and
is kept at $T_1$ after the aging at $T_2$ for a period of $\tau_2$, 
$\epsilon''$ goes back to a value at $T_1$, which deviates from 
the reference curve, and then $\epsilon''$ decreases with time during 
the aging  process at $T_1$.

To elucidate the aging dynamics at a temperature in the
TC mode more in detail, it is useful to investigate the difference between
$\epsilon''$ and $\epsilon''_{\rm ref}$ as we have already done in the
CR mode.
In this case, the reference values $\epsilon''_{\rm ref}(T_j,\omega)$ 
at $T_j$ ($j=1,2$) are subtracted from
the observed dielectric losses $\epsilon''$ at $T_j$, respectively. 
Here, we define $\epsilon''-\epsilon''_{\rm ref}\equiv\epsilon''(T,\omega
; t_{w1},t_{w2})-\epsilon''_{\rm ref}(T,\omega ; 0,0)$, where
$t_w$=$t_{w1}+t_{w2}$. For the TC mode,
\begin{widetext}
\begin{eqnarray}\label{eps2_age2}
\epsilon''-\epsilon''_{\rm ref}&\equiv& \epsilon''(T,\omega
; t_{w1},t_{w2})-\epsilon''_{\rm ref}(T,\omega ; 0,0)\nonumber\\
 &=& \left\{ \begin{array}{c@{\quad:\quad}l} 
\epsilon''(T_1,\omega ; t_w,0)-\epsilon''_{\rm ref}(T_1,\omega ; 0,0)  & 0<t_w<\tau_1 \\ 
\epsilon''(T_2,\omega ; \tau_1, t_w-\tau_1)-\epsilon''_{\rm ref}(T_2,\omega ; 0,0)  & \tau_1<t_w<\tau_1+\tau_2 \\ 
\epsilon''(T_1,\omega ; t_w-\tau_2, \tau_2)-\epsilon''_{\rm ref}(T_1,\omega ; 0,0)  & t_w > \tau_1+\tau_2  
\end{array}\right. 
\end{eqnarray}
\end{widetext}

Figure 7 displays the time
dependence of $\epsilon''-\epsilon''_{\rm ref}$ for the thin films
with $d$=20nm, $T_1$=375.3K, $T_2$=355.8K, and $\Delta T$=$-$19.5K.
The origin of the time ($t_w$=0) is defined as the time
when the temperature reaches $T_1$ after the cooling from 403 K (above
$T_g$). The value of $\epsilon''-\epsilon''_{\rm ref}$
is found to decrease with increasing $t_w$ and 
to change immediately to 0 when the
temperature is lowered from $T_1$ to $T_2$ at $t_w=\tau_1$.
This indicates that the decrease in temperature from $T_1$ to $T_2$ 
causes the polymer sample to {\it rejuvenate}. 
As the aging time elapses at $T_2$ after the temperature shift, 
$\epsilon''-\epsilon''_{\rm ref}$ 
is found to exhibit a relaxation just like a new relaxation process 
starts at $t_w=\tau_1$. 
The dotted curve in Fig.7(a) which starts at $t_w=\tau_1$ is the relaxation
curve of $\epsilon''-\epsilon''_{\rm ref}$ 
observed at $T_2$ after the temperature change directly from
403 K to $T_2$. The origin of the time axis is shifted from 0 to 
$t_w=\tau_1$ for the dotted curve in Fig.7(a).
The observed data for $\tau_1<t_w<\tau_1+\tau_2$ are located 
exactly upon this dotted curve. 
At $t_w=\tau_1+\tau_2$, the system is
heated up to $T_1$, and then the temperature of the system is kept at
$T_1$. On this stage, $\epsilon''-\epsilon''_{\rm ref}$ is found to 
go back to the
value which $\epsilon''-\epsilon''_{\rm ref}$ had reached at $t_w=\tau_1$
and to begin to decrease as if there were no temperature change
during the aging process at $T_1$.

In Fig.7(b) is shown the time evolution of 
$\epsilon''-\epsilon''_{\rm ref}$ after removing the data between
$\tau_1$ and $\tau_1+\tau_2$ and shifting the data for $t_w>\tau_1+\tau_2$
in the negative direction along the time axis by $\tau_2$. 
We can consider the horizontal axis as the total time which the
system spent at $T_1$. In this figure, we can see that 
$\epsilon''-\epsilon''_{\rm ref}$ decreases monotonically with the 
aging time 
without any discontinuous change except for a short region just after
$t_w=\tau_1$. The curve obtained in the above way agrees very well 
with that obtained by keeping the system at $T_1$ without any temperature
change, $\epsilon''(T_1,\omega ; t_w,0)-\epsilon''_{\rm ref}(T_1,\omega
; 0,0)$ (We call this curve {\it the standard relaxation curve}.). 
This implies that polymer glasses can remember the state at $t_w=\tau_1$
of the relaxation towards the equilibrium state and recall the memory at
$t_w=\tau_1+\tau_2$. We call this behavior a {\it complete
memory effect}.
This was a result obtained for $\Delta T=-19.5 K$.

\begin{figure*}
\includegraphics*[width=14cm]{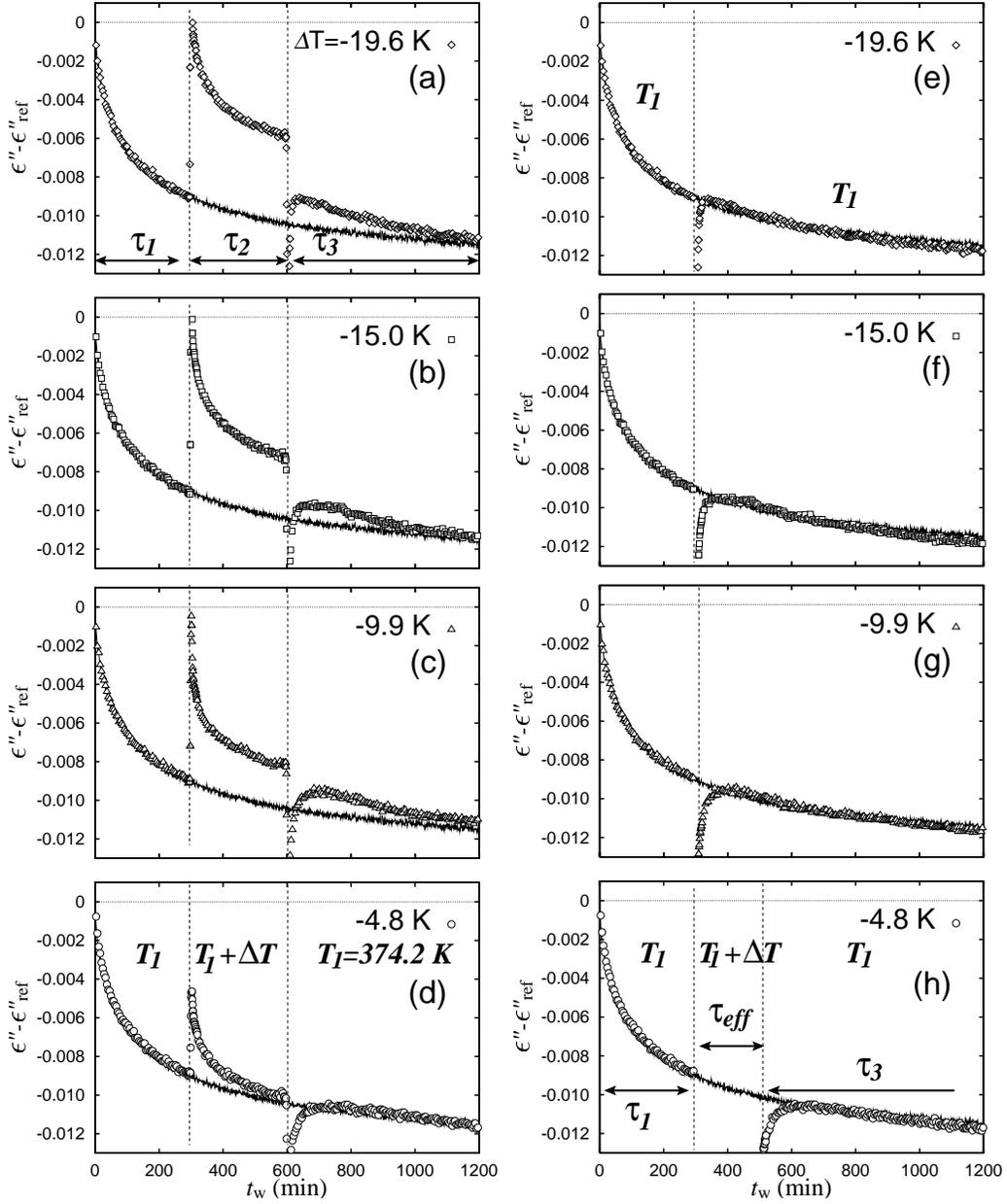}
\caption{\label{fig:8} Difference between $\epsilon''$ and
 $\epsilon''_{\rm ref}$ observed by the TC mode for PMMA thin films with  
$d$=26nm and $f$=20Hz. The aging temperature of the first stage $T_1$ is 
fixed to be 374.2K and $T_2(\equiv T_1+\Delta T)$ is changed: $\Delta
T$=$-$4.8 K (d,h), $-$9.9 K (c,g), $-$15.0 K (b,f), and $-$19.6 K (a,e).  
Aging times are $\tau_1=\tau_2=5$ hours. In Figs. 8(e)-(h), the 
data points of the second stages are removed
 and those of the third stage are shifted by an amount of time  in the 
negative direction of the time axis so that the data points of the third 
stages can be overlapped with the standard relaxation curve most reasonably.
The effective time $\tau_{\rm eff}$ is evaluated as the difference
between the time at the end of the first stage and the time at the
beginning of the initial time of the third stage after the
above data fitting process. The difference in length between
the arrow of $\tau_2$ in (a) and the arrow of $\tau_{\rm eff}$ 
in (h) shows how the aging at $T_2$ affects the aging at $T_1$.
}
\end{figure*}

\begin{figure*}
\includegraphics*[width=14cm]{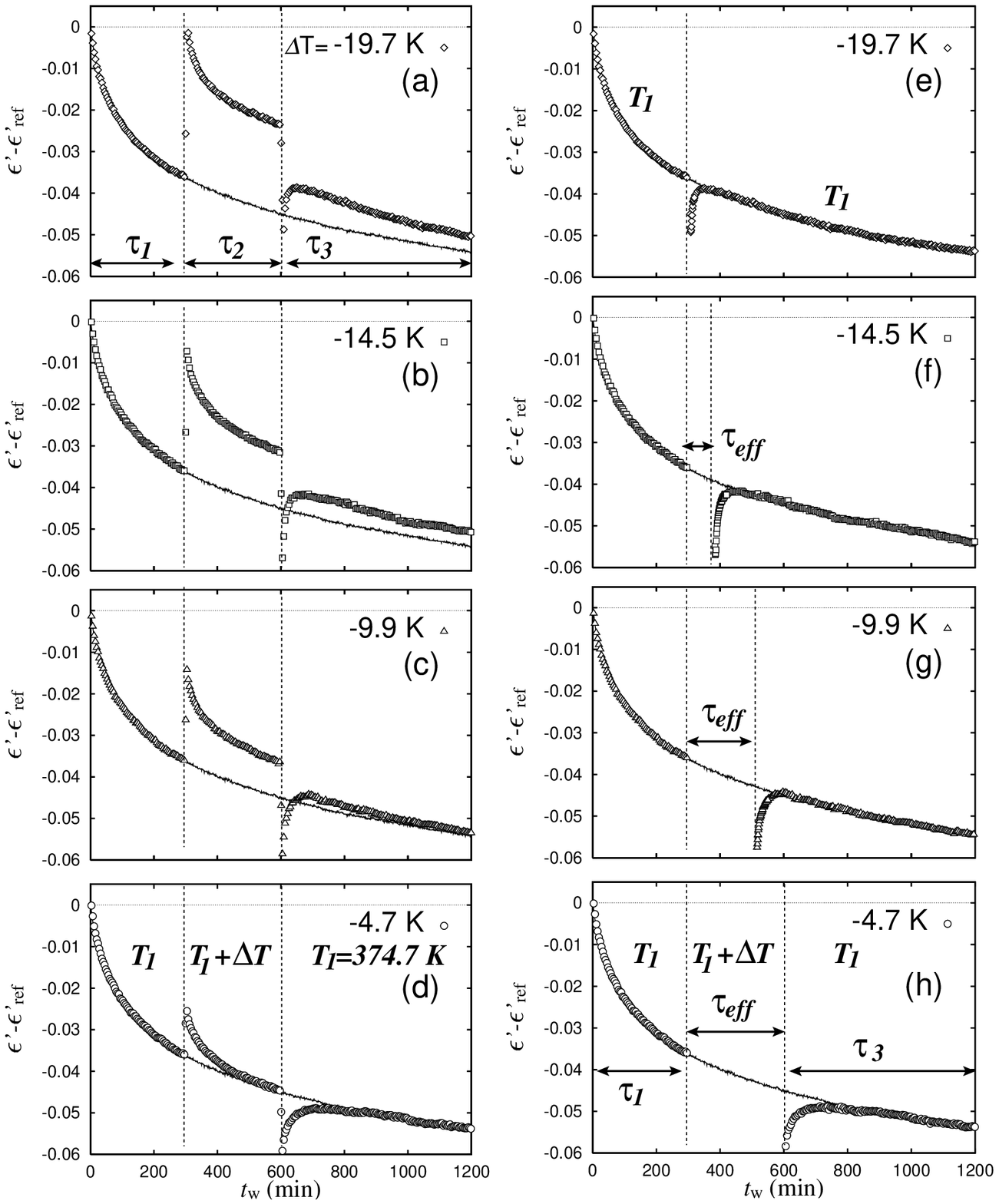}
\caption{\label{fig:9} Difference between $\epsilon'$ and
 $\epsilon'_{\rm ref}$ observed by the TC mode for PMMA films with
 $d$=514nm and $f$=100Hz. 
The aging temperature of the first stage $T_1$ is 
fixed to be 374.7K and $T_2(\equiv T_1+\Delta T)$ is changed: $\Delta
T$=$-$4.7 K (d,h), $-$9.9 K (c,g), $-$14.5 K (b,f), and $-$19.7 K (a,e).  
Aging times are
 $\tau_1=\tau_2=5$ hours. In the right figures, $\tau_{\rm eff}$ is 
the same as defined in Fig.8.
}
\end{figure*}

We here show the aging behavior for various values of $\Delta T$ 
for PMMA thin films with $d=26$nm.
Figures 8(a)-(d) show $\epsilon''-\epsilon''_{\rm ref}$ as a function of the
aging time $t_w$ for various values of $\Delta T$= $-$19.6 K to 
$-$4.8 K and $T_1=$374.2K. Figures 8(e)-(h) are obtained from 
Figs.8(a)-(d) according to the following procedure:
the data points for $t_w=\tau_1\sim\tau_1+\tau_2$ are removed and the ones
for $t_w>\tau_1+\tau_2$
are shifted by an amount of $\Delta\tau$ in the negative
direction along the
$t_w$ axis so that the data can be overlapped with the 
standard relaxation curve for the aging at $T_1$ in the most reasonable
way. Here, an effective time $\tau_{\rm eff}$ is defined as the time
$\tau_2-\Delta\tau$. The effective time can be considered as a measure
of the contributions of the aging at $T_2$ to that at $T_1$. If
$\tau_{\rm eff}$ is zero, the aging at $T_2$ has no contributions to
that at $T_1$, while if $\tau_{\rm eff}$ is equal to $\tau_2$, the aging
at $T_2$ has full contributions to that at $T_1$.
%
%
In Figs.8(a)-(d), it is found that the value of
$\epsilon''-\epsilon''_{\rm ref}$ is reinitialized at $t_w=\tau_1$ for
$\Delta T$=$-$19.6K$ \sim$ $-$9.9K. However, the reinitialization at
$t_w=\tau_1$ is not complete for $\Delta T$=$-$4.8 K.

As shown in Figs.8(e)-(h), $\tau_{\rm eff}$=0 for 
$\Delta T$=$-19.6$K$\sim$$-$9.9K, and $\tau_{\rm eff}\approx 200$min 
for $\Delta T$=$-4.8$ K.
This indicates that for $|\Delta T|\ge$9.9K there is a complete memory effect
and the aging at $T_2$ gives no contributions to that at $T_1$, in other
words, the aging at $T_1$ is independent of that at $T_2$. On the
contrary, for
$\Delta T$= $-4.8$ K, the aging at $T_2$ is no longer independent of that
at $T_1$, but gives some contributions to that at $T_1$. As a result,
$\tau_{\rm eff}$ has a positive finite value.
It should be noted here that, although 
the above result is obtained for $\epsilon''$, almost similar results
are obtained also for $\epsilon'$. 

Next, we show the result obtained for PMMA films with $d$=514 nm,
which is much larger than $d$=26 nm. As
an example, the observed results of $\epsilon'$ for $f$=100Hz are 
shown in Fig.9.
It is found from Figs.9(e)-(h) that for $\Delta T$=$-19.7$ K, 
$\tau_{\rm eff}$ is
zero, that is, the aging dynamics at $T_1$ are totally independent of
that at $T_2$. However, as $|\Delta T|$ decreases, $\tau_{\rm eff}$
increases monotonically. For $\Delta T$=$-4.7$ K, $\tau_{\rm
eff}\approx\tau_2$. This means that the aging at $T_2$ has almost full
contributions to that at $T_1$. On the basis of these results, it can 
be expected that the contributions of the aging at $T_2$ to that at
$T_1$ change continuously with decreasing $|\Delta T|$. 
A similar continuous change in $\tau_{\rm eff}$ with respect to 
$|\Delta T|$ has also been observed in some examples of spin
glasses~\cite{Sasaki1,Jonsson3}. Comparing Fig.8 with Fig.9, it is found
that thinner PMMA films are more sensitive to the change in
$|\Delta T|$ than thicker PMMA films.

To measure the contributions of the aging at $T_2$ to that at $T_1$, we
evaluate the ratio $\tau_{\rm eff}/\tau_2$ as functions of $\Delta
T$ under various conditions. In the case of $d=514$nm, we use 
the values of $\epsilon'$ for $f=100$Hz and 20Hz,
and those of $\epsilon''$ for $f=100$Hz, while in the case of $d=26$nm, 
we use the values of $\epsilon'$ for $f=20$Hz and those of $\epsilon''$ 
for $f=20$Hz. 
Figure 10 displays that the effective time $\tau_{\rm eff}$ normalized
with $\tau_2$ is a decreasing function of $|\Delta T|$ and 
decays to zero as $|\Delta T|$ increases.
The decaying rate of $\tau_{\rm eff}/\tau_2$ with respect to $|\Delta T|$
is larger for $d$=26nm than for $d$=514 nm. Therefore, we can say that 
the independence of the aging dynamics between two different temperatures
is accomplished faster in thinner films with increasing $|\Delta T|$.

\begin{figure}
\includegraphics*[width=8cm]{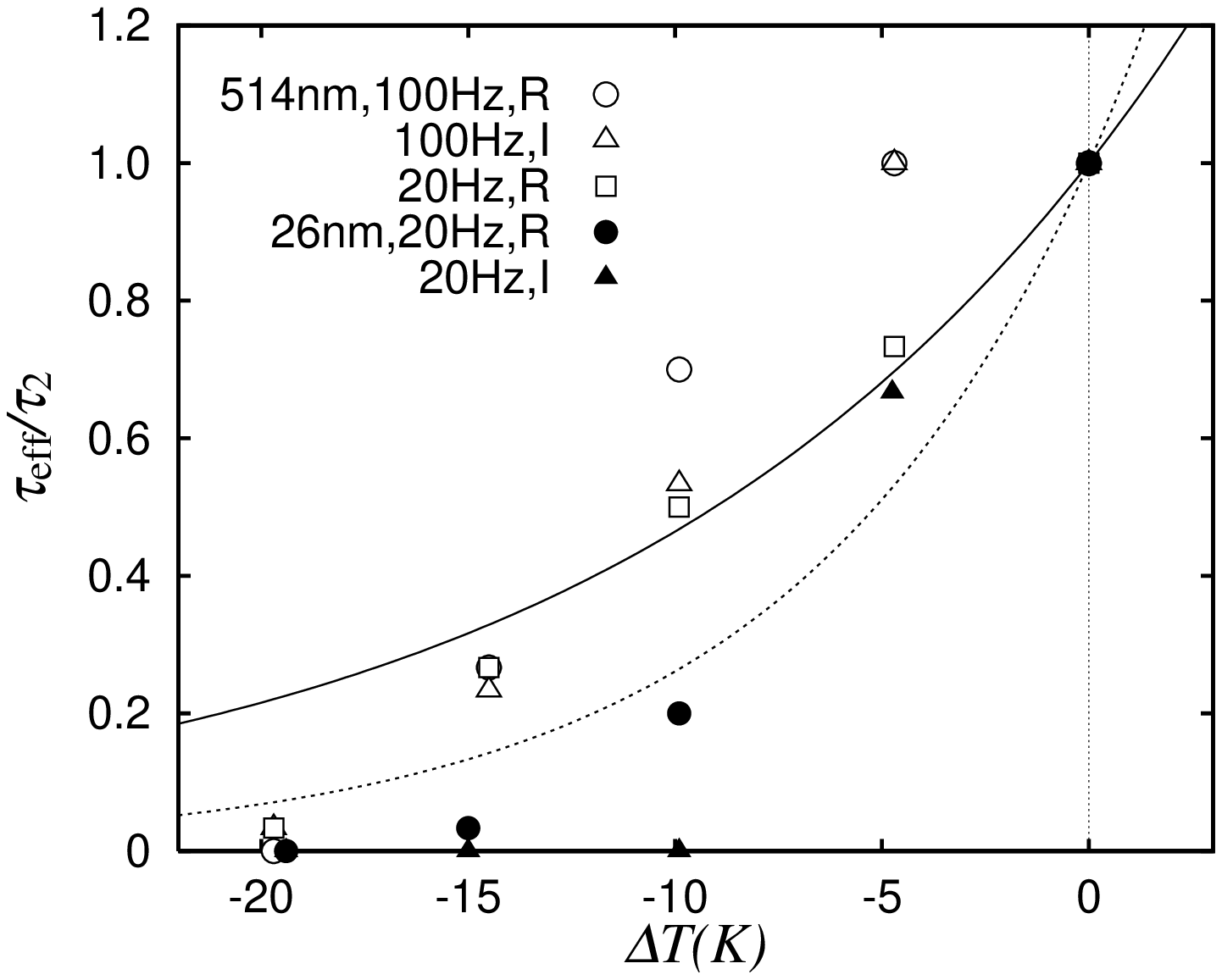}
\caption{\label{fig:10} Dependence of effective time $\tau_{\rm eff}$
 normalized with the aging time of the second stage $\tau_2$ on the
 difference of two aging temperatures for PMMA thin films with $d$=26nm,
 514nm and $f$=20Hz, 100Hz. The effective time $\tau_{\rm eff}$ is
 evaluated from the data for $\epsilon'$ and  $\epsilon''$. Open and full
 symbols display the results for $d$=514nm and $d$=26nm, respectively. 
The curves are obtained by fitting the data to Eq.(8). 
The solid curve is for $d=514$nm and the dotted curve is for $d=26$nm.
}
\end{figure}

It is interesting to compare the $\Delta T$ dependence of $\tau_{\rm
eff}$ with that estimated through a thermal activation process~\cite{Dupuis1}.
If it is assumed that the aging involves a thermally activated jump over a
free energy barrier $\Delta U$, the corresponding time
$\tau$ at temperature $T$ characteristic of the jump motion over 
the barrier is given by $\tau(T)=\tau_0\exp(\Delta U/k_BT)$,
where $\tau_0$ is a microscopic time scale and $k_B$ is the Boltzmann
constant. 
Here, we consider the aging at $T_2$ for a period of $\tau_2$.
In this case, we can say that the aging process proceeds by a 
reduced time $\tilde\tau$($\equiv \tau_2/\tau (T_2)$).
If the reduced time $\tilde\tau$ of a state is the same as that of
another state, we can consider that both the states 
are the same. Because the aging process at $T_2$ for 
a period of $\tau_2$ corresponds to that at $T_1$ for a period 
of $\tau_{\rm eff}$, we obtain the relation 
$\tau_{\rm eff}/\tau (T_1)=\tau_2/\tau (T_2)$. 
If the energy barrier $\Delta U$ is independent of $T$, we
obtain the relation 
$(\tau(T_1)/\tau_0)^{T_1}=(\tau(T_2)/\tau_0)^{T_2}$ 
for the thermal activation process.
Combining the above two equations, 
we obtain the relation for the temperature shift
from $T_2$(=$T_1+\Delta T$) to $T_1$:
\begin{eqnarray}
\frac{\tau_{\rm eff}}{\tau_2}=\left(\frac{\tau (T_2)}{\tau_0}\right)^{\Delta T/T_1}.
\end{eqnarray}
Because the determination of $\tau_{\rm eff}$ is accompanied with the
relatively large uncertainty, it is difficult to judge whether Eq.(8)
can reproduce the observed data in Fig.10. Nevertheless, from the
qualitative comparison of the observed results with Eq.(8), it is found
that the microscopic time, $\tau_0$, associated with the aging is smaller
in thin film geometry than in the bulk films.

\section{Discussions}
\subsection{Physical origin of $t_w$ dependence of $\epsilon'$ and
$\epsilon''$ }
As shown in Sec.III, we have observed the interesting behavior of 
the aging dynamics. In this section we discuss the physical origin of the 
$t_w$ dependence of $\epsilon'$ and $\epsilon''$ observed in the present
measurements.
The dynamic dielectric constant, 
$\epsilon^*(T)$, is obtained by dividing the complex electric 
capacitance $C^*(T)$ at the temperature $T$ by the geometrical 
capacitance $C_0(T_0)$ evaluated at $T_0$=273 K. 

The evaluation of $\epsilon^*$ in this paper is based on the assumption
that the $T$ and $t_w$ dependence of $C_0$ 
can be neglected compared with that of the {\it intrinsic} dielectric 
constant. 
This assumption may not be valid in some cases.
The $T$ dependence of $C_0$ can be neglected as long as 
the value normalized with that at $t_w$=0 is 
used for a given temperature. However, the $t_w$ dependence of $C_0$ 
cannot be neglected for the following two reasons.
Firstly, it is well-known that the physical aging of polymers 
increases the density
with aging time. This is called densification, which leads to the 
decrease in film thickness with increasing $t_w$. This 
contraction should be commonly observed independent of film thickness. In
addition to this,  
recent measurements done by Miyazaki {\it et~al.} show that the film 
thickness of polystyrene (PS) thin films 
supported on Si substrate decreases very slowly with time at an annealing
temperature above $T_g$ when the initial thickness is less than
20nm~\cite{Miyazaki1,Kanaya1}. This indicates that 
there should be a very slow 
relaxation process of PS in {\it thin film geometry}. 
If we take into account the above two processes for PMMA 
thin films, it follows that
the geometrical capacitance $C_0$ depends on the aging time $t_w$ 
and that the $t_w$ dependence of $C_0$ of ultra thin films  can be different
from that of bulk samples.

The effect due to the $t_w$ dependence of $C_0$ is evaluated here. 
Since $C'=\epsilon'\epsilon_0\frac{S}{d}$,
the relative change in $C'$ is given by
\begin{eqnarray}
\frac{\Delta C'}{C'}=\frac{\Delta\epsilon'}{\epsilon'}-\frac{\Delta d}{d},
\end{eqnarray}
where $S$ is assumed not to change with time, and the symbol $\Delta
X$($X$=$C'$, $\epsilon'$, $d$) is $\Delta X\equiv X(t_w)-X(0)$.
The discussions in this paragraph can also be valid for
$\epsilon''$ and $C''$.
If the sample goes towards an equilibrium state during the aging 
process at a temperature, the densification proceeds and the thermal
fluctuations are suppressed. The former leads to the decrease in 
$d$, while the latter leads to the decrease in $\epsilon'$.
In this case, it is reasonable to expect that $\Delta d/d$ and 
$\Delta\epsilon'/\epsilon'$ has the same form with respect to $t_w$.
Although  the aging strength may be modified, we can evaluate
the $t_w$ dependence of $\epsilon'$ through that of $C'$. 
In thin film geometry, however, the $t_w$ dependence of $\epsilon'$ 
obtained thus may be affected by a possible 
existence of the very slow relaxation process of the film thickness.


As shown in Fig.1(b), 
we observed that $\epsilon'(\equiv C'(T)/C'_0(T_0))$ {\it increases} with
$t_w$ after decaying up to about 11 h during the aging at $T_a=375.4$K 
for PMMA thin films with the initial thickness of 20nm. 
This increase may be due to the existence of the very slow relaxation
process of the thickness observed only in thin film
geometry~\cite{Kanaya1}. The characteristic time of this very slow
relaxation process should be longer than that of the relaxation of the 
{\it intrinsic} $\epsilon'$ ($\epsilon''$) and the contraction due to
the physical aging.

Now let us try to estimate the effect of the existence of this very slow
relaxation process on the present experimental results. Comparing the
results in Fig.1(a) and Fig.1(b), we find that
$\Delta\epsilon''(t_w)/\epsilon''(0)\sim 0.062$ and
$\Delta\epsilon'(t_w)/\epsilon'(0)\sim 0.0048$ for $t_w=11$h 
at $T_a=375.4$K. An analysis of $\Delta d/d$ in the present measurements
shows that $\Delta d(t_w)/d(0)\sim$ 0.0009 
for $t_w=11$h at $T_a=375.4$K. Therefore, even if the increase in 
$\epsilon'$ exists under some conditions, the effect of this increase 
on the dynamics of the intrinsic $\epsilon'$ is less than 20\% when
$t_w$ is limited to $t_w<11$h.  
It should be noted here that this effect can only be observed for 
$\epsilon'$ near $T_g$ and for ultrathin films ($d=20,26$nm).
On the basis of this discussion, we believe that the observed 
dynamics of $\epsilon'$ and $\epsilon''$ in the present measurements
can safely be regarded as those of the intrinsic dielectric constants 
for PMMA thin films, except for some special conditions mentioned above.

The discussions here are based on the existence of the very slow
relaxation mode in PMMA thin films. In order to check the validity
of this assumption, we should measure the frequency spectrum of
dynamics dielectric constant in the lower frequency region. 

\subsection{Comparison with the aging scenario in spin glasses}
In the CR and TC modes, we observed the memory and
rejuvenation effects. There are many common properties between polymer
glasses and spin glasses. For spin glass systems, the memory and
rejuvenation effects
have been discussed by two different models : one is a hierarchical
model~\cite{Lefloch,Vincent} 
and the other is a droplet model~\cite{Bray1,Fisher1,Fisher2,Fisher3}. 

The hierarchical model is based on the mean-field picture
originating from the Sherrington-Kirkpatrick model~\cite{SK1}. 
It is assumed that 
the spin glass (SG) phase is characterized with a multi valley structure of
the free energy surface at a given temperature.
At a temperature $T_1$, the system relaxes over the many valleys formed
at that temperature. When the system is cooled from $T_1$ to $T_1+\Delta
T$($\Delta T<0$), each valley of the free energy splits into new 
smaller subvalleys. If
$|\Delta T|$ is large enough, the energy barriers separating the initial
valleys are too high to go among different initial valleys within the
period of $\tau_2$ at $T_1+\Delta T$.  Only relaxations within the
initial valleys can be activated, and hence the occupation number of each
initial valleys holds during the aging at $T_1+\Delta T$. When the system is
heated again from $T_1+\Delta T$ to $T_1$, the small subvalleys merge
back into their ancestors. Therefore, the memory at the end of the first
isothermal aging process at $T_1$ can be memorized and recalled at
$\tau_1+\tau_2$. The rejuvenation effects can also be explained in a
similar way.

As for the droplet model, the growth of SG ordered domain is taken into 
account. In a spin glass system, the growth rate of the domain is very
slowed down because of the heterogeneous distribution of the
frustration.
Furthermore, the memory and rejuvenation effects can be explained within
the droplet theory using the {\it temperature chaos concept}, in which 
the spin configuration or the energy landscape can be changed globally even 
for any infinitesimal temperature change on a length scale larger than the 
{\it overlap length}. 

Recent studies based on the droplet model produced
more powerful scenarios such as the droplets-in-domain 
scenario~\cite{Takayama1} and the ghost domain
scenario~\cite{Yoshino1,Jonsson1}. Furthermore, disordered systems
showing the memory effects have been classified into 2 categories as
follows:
\begin{enumerate}
\item
the system showing strong rejuvenation effects and memory effects:
The {\it strong rejuvenation} is due to the temperature chaos
     effect. The relaxation associated with the strong rejuvenation
     after the temperature shift from $T_1$ to $T_2$ is the same as that
     observed after the direct cooling to the aging temperature $T_2$.
\item
the system showing no strong rejuvenation, but a transient relaxation
     and memory effects: In this case, the strength of the transient
     relaxation after the temperature shift by $\Delta T$ becomes
     stronger with increasing $|\Delta T|$. The aging proceeds
     cumulatively even if the temperature is shifted. 
\end{enumerate}
In some systems, both properties can be observed 
depending on the temperature shift $\Delta T$. 
For a sufficiently small $\Delta T$
the aging proceeds cumulatively. When $\Delta T$ becomes large, the
aging is no longer cumulative but the noncumulative and strong
rejuvenation is observed just after the temperature shift.

Polymeric systems have been so far classified into the type 2). 
Although the memory effects have been observed, it has been reported
in the literature 
that no strong rejuvenation occurs after the temperature 
shift~\cite{Yoshino1}. 
However, the present measurements show that, after the negative
temperature shift from $T_1$ to $T_2$ by $\Delta T\approx -20$K, the 
relaxation of $\epsilon''$ due to the aging is reinitialized, $i.e.$, goes 
back to the value on the reference curve measured during the
cooling process. The values of the $\epsilon''$, then,  
begin to decay at $T_2$ along the relaxation curve measured 
after the cooling directly to $T_2$, as shown in Fig.7. 
From this results, we believe that there is a strong rejuvenation in
the case of PMMA for $\Delta T\approx -20$K. 
Furthermore, as $\Delta T$ decreases from $-4.8$K to $-19.6$K in Fig.8,
the contributions of the aging at $T_2$ to that at $T_1$ decrease, and 
the effective time $\tau_{\rm eff}$ decreases to 0 with 
increasing $|\Delta T|$ in the negative temperature cycle.
This experimental results imply that there is a crossover from the
cumulative dynamics to the noncumulative ones as the value of 
$|\Delta T|$ increases.
In order to check whether there is really the crossover from cumulative
to noncumulative, $i.e.$, whether there is the temperature chaos effect
in polymeric systems, the `twin-temperature shift' experiments should be
performed, which have been developed in spin glasses~\cite{Jonsson2}.

Comparing polymeric systems with spin glasses, the characteristic time
of the microscopic flip motion of dipoles or spins can be expected to be 
much larger in polymeric systems than in spin
glasses. Therefore, polymeric systems may be a model system for
investigating the aging dynamics in the shorter time regime, since an  
experimental time window corresponds to shorter time scales in polymeric 
systems compared with atomic spin glasses. 
Superspin glasses (strongly interacting nanoparticle systems)~\cite{Jonsson3}
can also be regarded as a model system for the same purpose, 
because they have the longer microscopic flip time of a superspin than
atomic spin glasses have. The polymeric systems may have a still longer 
microscopic flip time, and hence we expect that the polymeric systems
can be better model systems.

\subsection{Thickness dependence of aging dynamics}
As for the thickness dependence of the aging phenomena, we could obtain 
several interesting results: 
1) the relaxation strength due to the aging decreases with decreasing film
thickness.
2) the width of the dip in $\epsilon'-\epsilon'_{\rm ref}$ 
during the cooling process in the CR mode becomes 
smaller with decreasing film thickness.
3) the effective time $\tau_{\rm eff}$ in the TC mode decays to 
zero more quickly in a thin
film geometry with increasing $|\Delta T|$ in the negative $\Delta T$
region than in bulk systems.

As mentioned in Sec.IV,
some of the above results relating to the thickness dependence may be
explained by assuming that the fundamental microscopic flip time
$\tau_0$ of dipoles becomes smaller with decreasing film thickness. 
As shown in Fig.5, the width of the dip observed during the cooling
process in the CR mode is $\sim 20$K for $d=514$nm and  
$\sim 10$K for $d=26$nm in the case of PMMA films. 
In (atomic) spin glasses, where $\tau_0$ is much smaller than in polymer
glasses, the width of the dip of $\epsilon'$ in the CR mode is $\sim
3$K~\cite{Jonsson3}, 
which is even smaller than the value for the thinnest film thickness.
The width of the dip observed 
in the CR mode does not depend on the cooling rate for the range of
cooling rate from 0.08K/min to 0.33K/min, although the depth of the
dip strongly depends on the cooling rate~\cite{Bellon1}.

\section{Summary}
In this paper, we have investigated the aging dynamics in thin films of
PMMA with $d=20$nm$\sim$514nm through dielectric measurements. The
results obtained in the present measurements are as follows.
\begin{enumerate}
\item
The relaxation of dielectric constants is observed during the aging
     process at an aging temperature $T_a$. The relaxation strength
     decreases with decreasing $T_a$ and with decreasing film
     thickness. In the films of $d=20$nm and 26nm, the increase in
     $\epsilon'$ has been observed at $T_a=375$K, which may be due to
     the existence of the ultra slow relaxation process of the film
     thickness observed recently through the X-ray reflectivity measurements.
\item
In the CR mode, the aging at $T_a$ by the way of cooling to room
     temperature can be memorized as the dip in
     $\epsilon'-\epsilon'_{\rm ref}$ curve around $T_a$. During the
     subsequent heating process the memory can be recalled. The width of
     the dip in $\epsilon'-\epsilon'_{\rm ref}$ decreases with
     decreasing film thickness.
\item
In the TC mode with negative $\Delta T$, a strong rejuvenation effect
     of the difference between $\epsilon''$ and $\epsilon''_{\rm ref}$
     has been observed just after the negative shift from $T_1$ to $T_2$
     by $\Delta T\approx -20$K. At the same time, a complete memory
     effect could also be observed when the temperature goes back from
     $T_2$ to $T_1$.
\item
As $|\Delta T|$ decreases in the negative $\Delta T$ region in the TC
     mode, 
     the contributions of the aging at $T_2$ to that at $T_1$ become larger
     $i.e.$, the effective time $\tau_{\rm eff}$, which is a measure of
     the contribution, increases from 0 to $\tau_2$. A continuous change
     in $\tau_{\rm eff}$ with $\Delta T$ could be observed, which
     reminds us of a crossover between the noncumulative aging 
     and the cumulative one. The dependence of $\tau_{\rm eff}$ 
     on $\Delta T$ depends on the film thickness.
\end{enumerate}

In this paper, we focused on the aging dynamics observed in dielectric
constants. However, similar aging phenomena can also be observed in 
heat capacity of PMMA observed by temperature
modulated DSC~\cite{Fukao2}. It can be expected that the aging
phenomena observed in this paper are common ones for
any susceptibility including dielectric constant, magnetic
susceptibility, and heat capacity. 

\section{Acknowledgments}
The authors appreciate H.Yoshino and H.Takayama for useful
discussions.
This work was supported by a Grant-in-Aid for Scientific Research
(B) (No. 16340122) from Japan Society for the Promotion of Science and
for Exploratory Research (No. 16654068) from the Ministry 
of Education, Culture, Sports, Science and Technology of Japan.


\end{document}